\documentclass[sigplan,screen,nonacm,manuscript,natbib=false]{acmart}
\pdfoutput=1

\RequirePackage[datamodel=acmdatamodel, style=acmauthoryear]{biblatex}
\usepackage{listings}
\usepackage{cleveref}
\usepackage{float}
\usepackage[hyperfirst=true]{glossaries}
\makeglossaries{}

\newacronym[description={Source Lines of Code ignoring whitespace and comments}]{sloc}{SLoC}{Source Lines of Code}
\newglossaryentry{arity}{name={arity},plural={arities},description={number of arguments taken by a function}}
\newacronym[description={C Intermediate Language~\cite{necula2002cil}}]{cil}{CIL}{C Intermediate Language}
\newglossaryentry{nakedp}{name={naked pointer},description={pointer allocated by C, outside the OCaml heap}}
\newacronym{FFI}{FFI}{Foreign Function Interface}
\newacronym{AST}{AST}{Abstract Syntax Tree}
\newacronym[]{GC}{GC}{Garbage Collector}

\setacronymstyle{footnote-desc}

\begin{document}

\title{Targeted Static Analysis for OCaml C Stubs}
\subtitle{Eliminating gremlins from the code}
\subtitlenote{a nod to Goblint, the static analysis framework that the tool uses. The tool currently has the less impressive name `lintcstubs'}

\author{Edwin T{\"o}r{\"o}k}
\email{edwin.torok@cloud.com}
\affiliation{%
	\department{XenServer}
	\institution{Cloud SG UK Ltd}
	\streetaddress{Building 101, Cambridge Science Park}
	\city{Cambridge}
	\country{United Kingdom}
	\postcode{CB4 0FY}
}

\begin{abstract}
	Migration to OCaml 5 requires updating a lot of C bindings due to the removal of \gls{nakedp} support.
	Writing OCaml user-defined primitives in C is a necessity, but is unsafe and error-prone.
	It does not benefit from either OCaml's or C's type checking, and existing C static analysers
	are not aware of the OCaml GC safety rules, and cannot infer them from existing macros alone.
	The alternative is automatically generating C stubs, which requires correctly managing value lifetimes.
  Having a static analyser for OCaml to C interfaces is useful outside the OCaml 5 porting effort too.

	After some motivating examples of real bugs in C bindings a static analyser is presented that
	finds these known classes of bugs.
  The tool works on the OCaml abstract parse and typed trees, and generates a header file and a caller model.
  Together with a simplified model of the OCaml runtime this is used as input to a static analysis framework, Goblint.
  An analysis is developed that tracks dereferences of OCaml values, and together with the existing framework reports
  incorrect dereferences.
  An example is shown how to extend the analysis to cover more safety properties.
  
  The tools and runtime models are generic and could be reused with other static analysis tools.
\end{abstract}


\maketitle

\section{Introduction}

\subsection{C bindings maintenance}

OCaml is designed as a memory-safe language, as long as unsafe features are avoided (like the \verb|Obj| module, C stubs, etc.).
However, most OCaml code does need to interact with C code: it is the API of the Operating System,
and there is a vast amount of existing code in various languages, where interfacing with C is the common denominator.
The required C code doesn't necessarily have to be in the same process, but for practical and performance reasons it often is.

This C code has been traditionally written by hand, and the rules it must follow are documented in the language manual~\cite{cffi}.
Compliance with these rules and internal consistency escape OCaml's and C's static type checking --- the resulting code is more unsafe than OCaml or C code alone.
However, it does not benefit from OCaml's type checking, and not even from C's type checking (everything is a \verb|value|).
C static analysers do not know about these rules either, and they cannot infer them purely from analysing the existing macros
and the source code of the OCaml runtime.

Prior to OCaml 5 C, pointers could be safely cast to OCaml values as long as minimum alignment requirements were met~\cite{ffipointers4}.
There was also a single global runtime lock: only one OCaml thread could execute at a time. 
This restriction didn't apply to C threads and C code: it was thus desirable to release the runtime lock whenever long-running C code was invoked,
such that this code could run in parallel with another OCaml thread~\cite{ffilock4}.

With the introduction of OCaml 5 these safety rules have changed:
\begin{description}
	\item[with the removal of \glspl{nakedp}] it is no longer possible to cast C pointers to an OCaml \verb|value|.
		A pointer needs to be wrapped or transformed using one of the techniques in the manual~\cite{ffipointers5}.

	\item[with the removal of the global runtime lock] the runtime lock is now per domain, introducing more race conditions for code using global variables~\cite{interactionffi,ffilock5}.
\end{description}

C bindings have to be reviewed and ported, which is a manual and error-prone process.
During the review one shouldn't assume that the existing C binding is correct!
It also requires deep expertise in both OCaml and C to fully understand the interaction of the code and the safety rules,
which reduces to a small intersection of people with both of these skills.
Even though the author thought he understood the required changes, one of the patches that added OCaml 5 support to a library
has actually introduced a race condition on OCaml 4.x too until the mistake was realized and corrected.
The manual approach doesn't scale to porting thousands of functions, hence an automated tool is needed.

Sometimes the C part of the OCaml bindings are updated by C developers (e.g.\ as part of a larger API refactoring)
without fully understanding the consequences on the correctness or safety of the OCaml code: both the OCaml and C code still compile
and are correct in isolation, but are not correct when considered together! (e.g.\ the number of arguments to a binding changes).
To avoid such bugs affecting a released version of the software a static analyser that runs in a project's CI can reject such changes early.

\subsection{Just enough static analysis}

Developing a targeted tool means it is available almost immediately when the OCaml 5 porting work is happening,
rather than years later when the porting work would already be complete.
A gradual approach was chosen: develop a tool that can detect the bugs the author has just fixed,
both to detect more instances of that bug in other bindings, and to prevent the bug from being reintroduced.
The tool still needs to be extensible enough and built on solid foundations, such that it can be extended to detect more bugs,
and eventually prove the correctness of the bindings.

For this reason the~\cite{Goblint2016} static analysis framework was chosen, itself based on~\cite{necula2002cil}.
A good overview of how to implement a Goblint analyser can be found in~\cite[5.2 Structure of Goblint]{apinis2014frameworks}.

The tool is useful both for detecting issues that affect OCaml 4.x and to assist porting to OCaml 5.

A tool that can only detect a single bug may sound limited at first, but it is somewhat generic than that, as it detects
a class of bugs: accessing OCaml values with the runtime/domain lock released.

\subsection{Limitations}\label{sec:lim1}

\verb|lintcstubs| cannot prove that bindings are fully correct, and doesn't yet support all the features of the C interface
such as callbacks, however it aims at a near zero false positive rate,
such that it can be used in a Continuous Integration pipeline to reliably detect the bugs it does know about.

Static analysis of a real-world language can run into scalability issues (e.g.\ deeply nested conditions will require approximation to avoid exponential state size),
and general issues of undecidability. This doesn't exclude successfully analysing simpler forms.
In practice, it is expected that the complexity of C stubs will be kept to a minimum, and most of the complicated
logic will be either in the caller (OCaml code), or the callee (the C library function we are wrapping with the stub).
In particular the logic of releasing and reacquiring the runtime lock is not expected to be conditional in itself,
it would be very difficult to reason about the correctness of such logic even for the C stub author. 
The analyser works on more than just toy examples: it is able to analyse the entirety of the C stubs used by the Xen and XAPI projects
(about 100 stubs).

Goblint can be tuned with different analysis time/precision trade-offs, e.g.\ it can use Apron~\cite{jeannet2009apron}
to increase the precision of its value range analysis, and various other analysis can be configured to be path-sensitive or not.

A fundamental limitation of this approach is that it can only see values, control flow and invariants on the C side,
and doesn't know about any invariants from the OCaml side.
This gap could be bridged in the future by implementing some minimal shape analysis of the OCaml types passed to and returned from
the C stubs, however full range analysis would only be possible if we perform static analysis on the OCaml code itself,
e.g.\ by compiling both languages using LLVM following ideas from Duplo~\cite{licker2020duplo}.

The approach described here is not limited to a single static analyser: the runtime model, and generated primitive models
could also be used with other static analysis frameworks in the future, e.g.\ by symbolically executing
the generated model and real C code using~\cite{klee}. Or by generating pre- and post-condition validation code that would complement
the runtime assertions generated by~\cite{filliatre2021ortac}.

Alternatives are described in~\cref{section:related}.

\section{Arity mismatch}\label{bug:arity}

\Cref{lst:origarity1,lst:origarity2}~define an OCaml primitive \verb|domain_assign_device|. The primitive is implemented in a C function
\verb|stub_xc_domain_assign_device|.

\begin{lstlisting}[caption={Original OCaml primitive declaration},label=lst:origarity1,language={[Objective]Caml}]
external domain_assign_device: handle -> domid -> (int * int * int * int) -> unit
  = "stub_xc_domain_assign_device"
\end{lstlisting}

\begin{lstlisting}[caption={Original C stub implementation},label=lst:origarity2,language={C}]
CAMLprim value stub_xc_domain_assign_device(value xch, value domid, value desc)
{
  /* ... */
}
\end{lstlisting}

This is correct: the OCaml function takes 3 parameters and returns \verb|unit|, and the C function takes 3 parameters too.
Below 5 parameters bytecode and native code primitive implementations are identical, so a single stub is correct.
Above 5 parameters different stubs need to be provided for bytecode and native code~\cite{arity5}.

\Cref{lst:aritylisting}~is a change in the C stubs that looks correct on its own, it introduces a new parameter:

\begin{lstlisting}[caption={Refactoring introduced new parameter},label=lst:aritylisting,language={[Objective]Caml}]
-CAMLprim value stub_xc_domain_assign_device(value xch, value domid, value desc)
+CAMLprim value stub_xc_domain_assign_device(value xch, value domid, value desc,
+                                            value rflag)
\end{lstlisting}

It introduces a bug. Notice what is missing from the patch~\cref{lst:origarity1}:
an update to the \verb|.ml| file, which declares a function with just 3 arguments.

This change may seem correct to a C developer, and expect the compiler to either raise an error
when it detects the argument mismatch, or show a warning if calling the function without having a prototype declared.
The OCaml compiler can do that too for pure OCaml code, but when combining C and OCaml neither compiler is aware of the type,
or even the \gls{arity} of the function in the other language.

\section{Arity and C argument type checking}\label{sec:arity}

Checking whether the arguments for the C \gls{FFI} are correct in general would require static analysis
(shape analysis or a specific type system as in~\cite{furr2008checking}).
However, in the particular case of~\cref{bug:arity} we can detect the \gls{arity} mismatch at compile time with a simple tool~\cite{lintcstubs-arity}:
generate a C header file from the OCaml \verb|.ml| file automatically, and let the C compiler check the \gls{arity}.

It might be useful to eventually ship such a tool as part of the OCaml compiler, but we don't have to wait for that
thanks to~\cite[compiler-libs]{compilerlibs}, which allows to implement a simple OCaml \gls{AST} processor for generating the header file
in less than 50 \gls{sloc} for checking the correctness of bytecode stubs with no external dependencies.
This is a tool that produces a \verb|.h| file when given a \verb|.ml| file as input.
It is suitable to be run as part of the Xen build process, which is quite stringent about external dependencies,
and the build system has to run on old systems, such as Debian oldstable.

Native stubs may use more advanced features (e.g.\ unboxing~\cite{cheaperc}). Checking their correctness requires a more advanced form of \gls{arity} checking based on the OCaml typed tree,
because we need to handle unboxing, etc.\ and the typed tree contains all that information already without
having to duplicate processing of the attributes in the static analyser
(see \lstinline|Val_prim| and the \lstinline|Primitive| in~\cite{compilerlibs}).
It should also keep up with new features in the C FFI (attributes) reasonably well without having to update the static analyser
on each compiler release.

To get a typed tree from a \verb|.ml| file requires building files in dependency order with appropriate build flags (e.g.\ include directories).
The build system has already performed that work, and with the \verb|-bin-annot| flag
the OCaml compiler will dump the typed tree it produces during compilation to \verb|.cmt| and \verb|.cmti| files.
These files are also used as input by documentation generation tools.
This tool takes a \verb|.cmt| file as input and produces a \verb|.h| file.
It can be run as part of a CI and is only slightly larger than the previous tool at $\approx 100$ \gls{sloc}.
If only a \verb|.cmti| file is available the tool can use that, but it won't necessarily find all the primitives,
because the \verb|.mli| is not guaranteed to contain them.

It checks that we have the correct number of arguments, and the type for unboxed arguments and return types.

Because this works on already compiled \verb|.cmt*| files we don't need to duplicate build arguments like include paths,
and can be integrated quite easily in \verb|dune| as an additional rule that takes \verb|%{cmti:filename}| as input.

Now that we have a way to process primitives and their \glspl{arity} and C argument types we can use this information
to generate another file that would be useful for a static analyser: a \lstinline|main| function that invokes all
C primitives with auto-generated values. This can be useful for reachability analysis, and to avoid false positives about
\lstinline|NULL| pointer dereferences (an OCaml \lstinline|value| is never \lstinline|NULL|, but a generic static analyser won't know that).
The invocations are placed in a new pthread to show the static analyser that these functions can be called in parallel
and with the runtime or domain lock held as appropriate for the OCaml major version.
This takes up another $\approx 150$ \gls{sloc} in a tool called \verb|lintcstubs_genmain|.

\begin{figure}[H]
\centering
\includegraphics[width=0.9\linewidth]{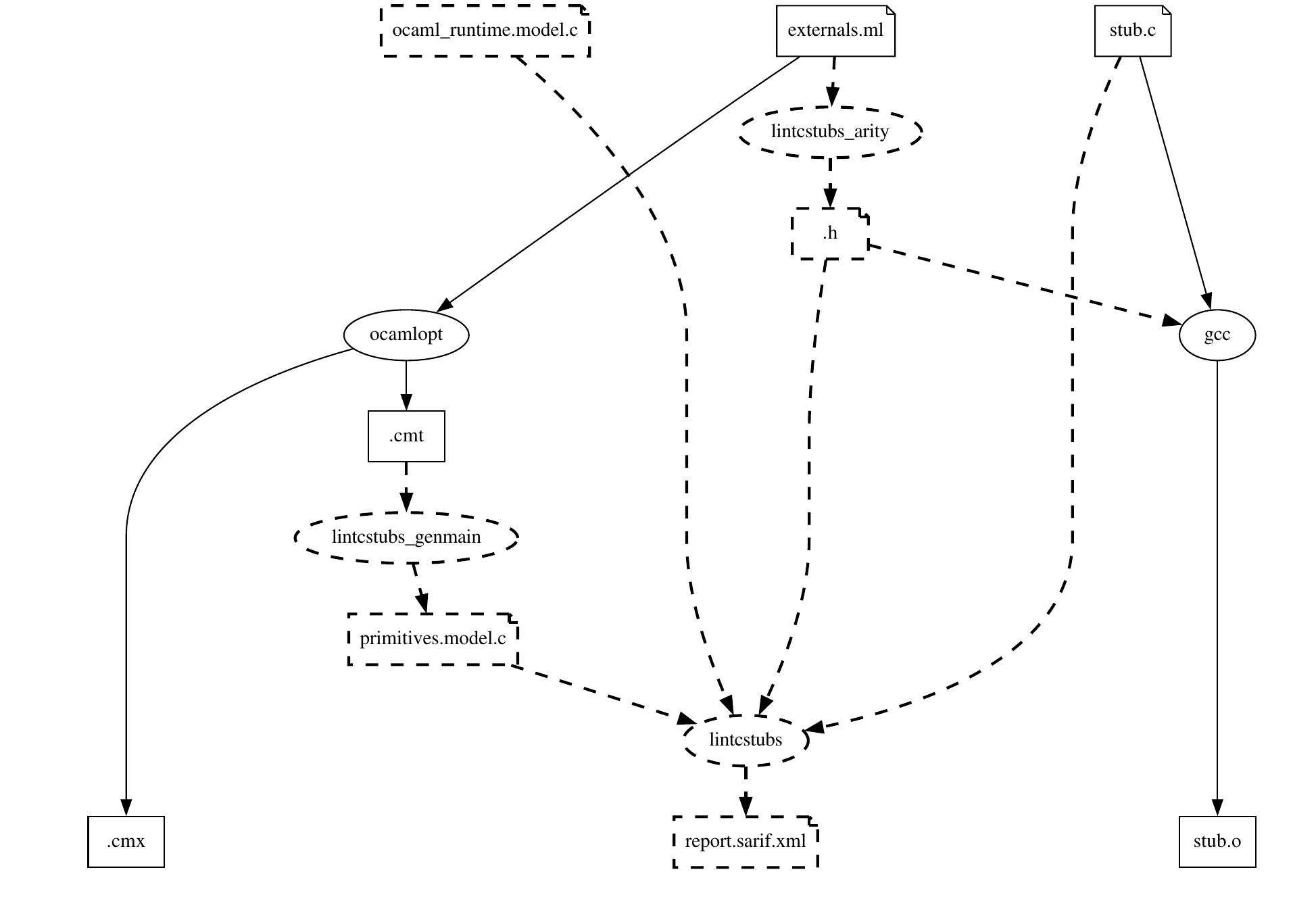}
\Description{A diagram showing the inputs, intermediate files, outputs and how the lintcstubs tool fits in}
\caption{Build flow diagram}\label{fig:toolflow}
\end{figure}

The build flow with these tools is shown in~\cref{fig:toolflow}:
\begin{itemize}
  \item \verb|lintcstubs_arity| generates a header file out of the C primitive declarations in a \verb|.ml| file
  \item \verb|lintcstubs_genmain| generates a \verb|.c| file containing a \verb|main| function that invokes all C stubs
  \item \verb|lintcstubs| is the goblint based static analyser described in \cref{sec:analyzer}
  \item \verb|ocaml_runtime.model.c| is a simplified model of the OCaml runtime written by hand
\end{itemize}

To test that the \gls{arity} checker works \verb|dune|'s \verb|cram|-style test feature is used in~\cref{lst:cram} to exercise some well known corner cases.

\begin{lstlisting}[caption={CRAM-style test},label=lst:cram]
Arity > 5 is implemented differently in bytecode and native code:
  $ cat >test.ml <<EOF
  > external add_nat: nat -> int -> int -> nat -> int -> int -> int -> int
  >                 = "add_nat_bytecode" "add_nat_native"
  > EOF
  $ lintcstubs_arity test.ml
  /* AUTOGENERATED FILE, DO NOT EDIT */
  #define CAML_NAME_SPACE
  #define _GNU_SOURCE
  #define _XOPEN_SOURCE 600
  #include <caml/mlvalues.h>
  CAMLprim value add_nat_bytecode(value *argv, int argn);
\end{lstlisting}

A perhaps surprising bug was discovered where C bindings that take no argument were declared as \lstinline| (void)|,
but the OCaml primitive will always pass a \verb|unit| (integer 0) value.
In practice calling conventions usually pass this value in a register and this mismatch may be harmless,
however for consistency and portability the C stub should still be fixed to always take at least one argument.

\section{Race condition on Data\_custom\_val dereference}\label{bug:deref}

\Cref{lst:custom1} defines a primitive that allocates a C type \verb|xenevtchn_handle| at~\cref{line:evalloc},
casts it to an OCaml value at~\cref{line:evcast} and returns it as an OCaml type at~\cref{line:evret}.

\begin{lstlisting}[caption={first xeneventchn binding},escapechar=@,label={lst:custom1},language={C}]
/* cast OCaml 'value' to C pointer */
#define _H(__h) ((xenevtchn_handle *)(__h))@\label{line:Hdecl}@

CAMLprim value stub_eventchn_init(void)
{
        CAMLparam0();
        CAMLlocal1(result);
        xenevtchn_handle *xce;

        caml_enter_blocking_section(); /* releases the OCaml runtime/domain lock */
        xce = xenevtchn_open(NULL, 0); /* allocates a C object */@\label{line:evalloc}@
        caml_leave_blocking_section(); /* reacquires the OCaml runtime/domain lock */

        if (xce == NULL)
                caml_failwith("open failed");

        result = (value)xce; /* casts a C pointer to an OCaml 'value'*/@\label{line:evcast}@
        CAMLreturn(result);@\label{line:evret}@
}
\end{lstlisting}

In~\cref{lst:custom11} at~\cref{line:evnotify} is another primitive that casts the OCaml value back at~\cref{line:evcast2} using the \verb|_H| macro
declared at~\cref{line:Hdecl}.

\begin{minipage}{0.95\linewidth}
\begin{lstlisting}[caption={second xeneventchn binding},escapechar=@,label={lst:custom11},language={C}]
CAMLprim value stub_eventchn_notify(value xce, value port)@\label{line:evnotify}@
{
        CAMLparam2(xce, port);
        int rc;

        caml_enter_blocking_section(); /* releases the OCaml runtime/domain lock */

        /* casts the OCaml value to a C pointer and calls a C function */
        rc = xenevtchn_notify(_H(xce), Int_val(port)); @\label{line:evcast2}@

        caml_leave_blocking_section(); /* reacquires the OCaml runtime/domain lock */

        if (rc == -1)
                caml_failwith("evtchn notify failed");

        CAMLreturn(Val_unit);
}
\end{lstlisting}
\end{minipage}

On the OCaml side in~\cref{lst:custom1ml} the type is declared as the abstract type \verb|handle|.

\begin{lstlisting}[caption={xeneventchn OCaml interface},label={lst:custom1ml},language={[Objective]Caml}]
type handle
val init: unit -> handle (* allocate an event channel [handle] *)
val notify : handle -> t -> unit (* notify the event channel on given port *)
\end{lstlisting}

This is all valid OCaml code prior to version 5: the value here is a word-aligned C pointer.
C pointers are not moved by the OCaml \gls{GC}, and can be used just as a C pointer normally would.

OCaml 5 no longer supports passing C pointers as OCaml values, so-called ``\glspl{nakedp}''.
All pointers must either be wrapped in an outer OCaml value, or transformed to an integer and back using bit manipulation.
The former was chosen when converting this code.

The fix itself can be seen in~\cref{lst:custom} and looks correct on its own: it replaces the use of \gls{nakedp}
with a custom value wrapping it, and if this is all you see in a pull request you might be tempted to approve it:
\begin{lstlisting}[caption={Removing naked pointers},label=lst:custom,language={C}]
-/* cast OCaml 'value' to C pointer */
-#define _H(__h) ((xenevtchn_handle *)(__h))
-
+/* dereference the OCaml value, read C pointer contained inside */
+#define _H(__h) (*((xenevtchn_handle **)Data_custom_val(__h)))
+
/* ... */
-   result = (value)xce; /* cast OCaml value to C pointer */
+   /* allocate an custom OCaml block */
+   result = caml_alloc_custom(&xenevtchn_ops, sizeof(xce), 0, 1);
+   _H(result) = xce; /* store C pointer inside OCaml custom block */

\end{lstlisting}

The bug is in the part that did not change~\cref{lst:deref,line:custombug}:

\begin{minipage}{0.95\linewidth}
\begin{lstlisting}[caption={Dereferencing custom value without runtime lock},label=lst:deref,escapechar=@,language={C}]
    caml_enter_blocking_section(); /* release OCaml runtime/domain lock */

    /* BUG: dereference OCaml value and read C pointer */@\label{line:custombug}@
    result = xc_domain_create(_H(xch), &domid, &cfg); 

    caml_leave_blocking_section(); /* release OCaml runtime/domain lock */
\end{lstlisting}
\end{minipage}

Previously this was safe because \lstinline|_H| was just a type cast of a C pointer
(which by definition doesn't move), so it can be used anywhere.
However now the macro is no longer a typecast: it dereferences an OCaml value. OCaml values can be moved by the Garbage Collector
(at least on OCaml prior to version 5.0).

To make this patch safe, it is best to delete the macro completely, which ensures that
any attempt at using in a stale piece of code will result in a compile-time error.
(This is especially important because Xen has a patchqueue in XenServer, and that wouldn't necessarily
be on all reviewers minds or their responsibility to check).
Once deleted, replace it with a static function for converting the type as recommended in the manual,
and store it in a local variable~\cref{lst:fixderef,line:fixderef,line:fixderef2}.
This is safe, because the C pointer will not move.

This update is tedious but can be mechanized, and is fail-safe: if you don't update all places in the code it won't compile.

\begin{minipage}{0.95\textwidth}
\begin{lstlisting}[caption={Fixing the dereference bug},label=lst:fixderef,escapechar=@,language={C}]
-#define _H(__h) (*((xc_interface **)Data_custom_val(__h)))
+static inline xc_interface *xch_of_val(value v)
+{ /* wrapper to dereference an OCaml value and read out the C pointer */
+  xc_interface *xch = *(xc_interface **)Data_custom_val(v);
+  return xch;
+}

-CAMLprim value stub_xc_domain_create(value xch, value wanted_domid, value config)
+CAMLprim value stub_xc_domain_create(value xch_val, value wanted_domid, value config)
 {
-   CAMLparam3(xch, wanted_domid, config);
+   CAMLparam3(xch_val, wanted_domid, config); /* renamed to 'xch_val' */
    CAMLlocal2(l, arch_domconfig);
+   /* dereference OCaml value while holding the runtime lock */
+   xc_interface *xch = xch_of_val(xch_val);@\label{line:fixderef}@

    caml_enter_blocking_section();
-   result = xc_domain_create(_H(xch), &domid, &cfg);
+   /* correct: no OCaml value is dereferenced */
+   result = xc_domain_create(xch, &domid, &cfg);@\label{line:fixderef2}@
    caml_leave_blocking_section();

    if (result < 0)
-       failwith_xc(_H(xch));
+       failwith_xc(xch);
\end{lstlisting}
\end{minipage}

\section{Safety properties when the runtime lock is released}\label{sec:runtimelock}

The state of the runtime lock is tracked by the Goblint framework, \verb|lintcstubs| queries its \verb|MustBeProtectedBy| set.
When calls are made to known functions that are part of the OCaml runtime it is asserted that the runtime lock is held
by checking the membership of the OCaml runtime lock in this set (and a failure will generate a warning from the static analyser).
Note that membership can have three states: known to be always protected by the lock, known to be never protected by the lock,
and unknown (sometimes protected, sometimes not, or loss of precision during analysis).
Because we do not expect loss of precision here (the C stubs are usually fairly simple in their logic around acquiring/releasing the runtime lock),
we require the lock to be always known to be acquired, and show a warning in the other two cases.

We could just let the static analyser figure this out on its own, however the analyser by default will only see one side
of the race condition: the one in the C stub, and it wouldn't know that after releasing the runtime lock
the GC may be invoked that may move the values (or otherwise change them).
The manually written model for the OCaml runtime does attempt to guide the static analyser in this direction
by immediately calling \verb|__may_call_gc| when releasing the runtime lock, and this function simulates the \gls{GC} 
moving all the values. More work is needed to make this step more precise, hence teaching the analyser about the runtime lock
specifically increases the precision of the analysis and the reported error message.

Upon entry to a C stub the analyser assumes the runtime lock is held (and this would be checked if the stubs call each-other).
With the runtime lock released a diagnostic is generated if any calls are made to runtime functions or if OCaml value derived pointers
are dereferenced.

The accuracy of this diagnostics is limited by the accuracy of tracking the runtime lock state.
Works reasonably well for simple bindings that don't have a lot of conditional code (such as the Xen bindings),
but may not work in general.

\section{C pointer to Abstract\_val can become stale}\label{bug:abstractref}

Looking at~\cref{lst:abstractval} it seems correct at first glance: there is no dereference of an OCaml value,
\verb|intf| is a C pointer of type \verb|struct mmap_interface*|:

\begin{lstlisting}[caption={Stale pointer to OCaml value},label=lst:abstractval,escapechar=@,language={C}]
 /* allocate memory for a C structure, wrap it in an abstract OCaml value */
 result = caml_alloc(Wsize_bsize(sizeof(struct mmap_interface)),
                Abstract_tag);

 intf = (struct mmap_interface *) result;/* the C pointer now points inside an OCaml value! */@\label{line:intf}@
 intf->len = Int_val(size);

 caml_enter_blocking_section(); /* release OCaml runtime/domain lock */
 /* 'intf' points inside an OCaml value, 
     it is not safe to dereference this when not holding the runtime lock,
     because OCaml code that runs in parallel with this may invoke the GC,
     which could move the OCaml value.
     This would leave 'intf' pointing to stale memory.
 */
 intf->addr /* BUG: unsafe 'intf' deref */ = xc_map_foreign_range(xch_val, c_dom,
                                   intf->len, PROT_READ|PROT_WRITE,/* BUG: unsafe 'intf' deref */
                                   c_mfn);
 caml_leave_blocking_section(); /* reacquire OCaml runtime/domain lock */
\end{lstlisting}

Releasing the runtime lock in~\cref{lst:abstractval} is desirable, since mapping pages can be time-consuming.
However, the C pointer \lstinline|intf| in~\cref{line:intf} is just a typecast of the OCaml value
(shown here after expanding \verb|Data_abstract_val|), and with the runtime lock released
that can move, leaving \verb|intf| pointing to a stale area of memory (perhaps reused by another value or unmapped).

This is another instance of the bug described in~\cref{bug:deref}, except the pointer to the OCaml value has been cast to a C pointer and stored in a temporary variable.

The correct solution here is shown in~\cref{lst:fixabstract}: store all the values needed inside the blocking section into C variables~(the \verb|ptr| variable at \cref{line:ptr}),
and store them into the abstract value at the very end once the runtime lock is held again~(\cref{line:heldstore}).

This is not yet tracked by the static analyser at the time of this submission,
but highlights an important property: C pointers derived from OCaml values should be treated as unsafe for dereferencing
when the OCaml runtime lock is released, just as OCaml \lstinline|value| are.

\begin{minipage}{0.95\textwidth}
\begin{lstlisting}[caption={Fix race by never storing C pointers to Abstract\_val},label=lst:fixabstract,escapechar=@,language={C}]
 /* allocate memory for a C structure, wrap it in an abstract OCaml value */
 result = caml_alloc(Wsize_bsize(sizeof(struct mmap_interface)),
                Abstract_tag);

 caml_enter_blocking_section(); /* release OCaml runtime/domain lock */
 /* correct: store result in temporary C variable */
 ptr = xc_map_foreign_range(xch, Int_val(dom), len,PROT_READ|PROT_WRITE, c_mfn);@\label{line:ptr}@
 caml_leave_blocking_section();/* reacquire OCaml runtime/domain lock */
 if (!ptr)
     caml_failwith("xc_map_foreign_range error");

 /* correct: points inside the OCaml value, with runtime lock is held */
 intf = Data_abstract_val(result);

 /* correct: store data in abstract OCaml value with runtime lock held */
 *intf = (struct mmap_interface){ ptr, len };@\label{line:heldstore}@
\end{lstlisting}
\end{minipage}

\subsection{Storing C pointers to Abstract\_val on the stack}

The \gls{GC} may move the \verb|Abstract_val|, invalidating any C pointers to it.
C pointers that point inside OCaml values should never be stored in variables that live across enter/leave blocking section calls.

\subsection{Basics about lattices and the Goblint framework}

\Gls{cil} is a simplified subset of C for program analysis and transformation, with tools that can
parse a regular \verb|.c| file, and produce this simplified form, an OCaml AST representation,
and a static analysis framework to reason about properties of the program.
Goblint-CIL updates \gls{cil} to work with newer C standards (C11) and OCaml versions.
Goblint is a static analysis framework that builds on top of Goblint-CIL focused on analysing multithreaded programs.

The interested reader can learn more from~\cite{Goblint2016,apinis2014frameworks,ciltut}.

\section{Modelling safety properties}\label{sec:analyzer}

The OCaml runtime lock is modelled as a global pthread mutex \lstinline|__VERIFIER_ocaml_runtime_lock|,
which gets acquired and released by the OCaml runtime model and the main function generated by \verb|lintcstubs_genmain|.
The existing mutex state analysis from Goblint~\cite{apinis2014frameworks} is used to track the current locked or unlocked
state throughout the C stubs. There is a helper \lstinline|DomainLock| module that provides a
\lstinline|must_be_held| and \lstinline|must_be_protected_by| to be used by other analyses,
implemented by querying Goblint's must-lock set and must-be-protected-by lattice.
It issues a diagnostic error if the required conditions are not met with the name of the called function
and the source code location of the function call.
It currently considers a single global lock (as in OCaml <5.0), but eventually it should
be configurable to simulate N threads M domains instead.

There is a \lstinline|Cstub.call_caml_runtime| helper which uses the DomainLock module
to assert that the domainlock must be held, and issue a diagnostic warning otherwise.

A \gls{cil} visitor is implemented that detects whether dereferences in expressions
originate in OCaml value types.
This relies on OCaml values being declared with the \lstinline|value| type,
and no copies stored in other variables.
A more sophisticated points-to analyses may be used in the future to increase accuracy here.
The visitor is called from Goblint's \lstinline|Events.Access| events on known reachable value reads and writes.

An OCaml runtime model with malloc/free was implemented to make it easier for the static analyser to ``see''
the control and data flow. In particular the OCaml runtime has a lot of conditional code
to decide whether \lstinline|mmap| or \lstinline|malloc| is used, and we cannot expect
the static analyser to infer from the garbage collector when finalizer functions may be called.
To increase accuracy we model the OCaml runtime instead in C using the analyser's \lstinline|nondet| feature:
a call to a C function that non-deterministically returns an integer, and splits the static analyser's path-sensitive analysis
to consider all cases in its dataflow analysis.

We define a \lstinline|maybe_call_gc| in the model that invokes all finalizers non-deterministically,
and place calls to this function at all points where in a real program it would be expected
that the garbage collector may be called.

Runtime lock safety is not the only rule that is implemented, there is also an implementation for Rule 5.1: ``a function that has parameters or local variables of type value must begin with a call to one of the CAMLparam macros''.
(although this may have false positives for C stubs that do not allocate).

There are also some exceptions defined, e.g.\ \lstinline|caml_stat_free| doesn't require the
runtime lock to be held, although in general any function that starts with \lstinline|caml_| is required
to be called with the runtime lock held.

\lstinline|caml_alloc_custom| also needs special handling: it uses a points-to analysis
to determine all the function pointer targets from the custom operations and ensure
that the analyser knows that the finalizer may be called wherever the \gls{GC} is allowed to run.
(In fact a ``\gls{GC} may run here'' point is inserted immediately after releasing the runtime lock to catch most race conditions).

\subsection{Callbacks}

Callback from C to OCaml are notoriously difficult to model correctly. And in fact prior to OCaml 5.0 cannot be used correctly, unless
it is known that either all callers release the runtime lock, or none of them do.
Which is difficult to know if the binding is a generic logger callback that may be used in both contexts.
This requires whole program (or at least multimodule) dataflow to determine safety and is not currently implemented.

\subsection{Extending the tool}

The tool is meant to be extensible, and to show that on a concrete example we'll extend it to support detecting a new bug type.
The following code snippet is taken from the Xen codebase: \cref{lst:physinfo,lst:tagcons}.

\begin{minipage}{0.95\textwidth}
\begin{lstlisting}[caption={OCaml type definition},label=lst:physinfo,language={[Objective]Caml}]
type arm_physinfo_caps =
  { sve_vl: int; }
type x86_physinfo_cap_flag (* abstract type, currently unused *)
type arch_physinfo_cap_flags =
  | ARM of arm_physinfo_caps
  | X86 of x86_physinfo_cap_flag list (* a list with abstract type elements *)
\end{lstlisting}
\end{minipage}

\begin{minipage}{0.95\textwidth}
\begin{lstlisting}[caption={Using Tag\_cons instead of Val\_emptylist},label=lst:tagcons,escapechar=@,language={C}]
CAMLlocal2(arch_cap_flags, arch_obj);
tag = 1; /* tag x86 */

arch_obj = Tag_cons;/* BUG: tag stored into OCaml value */@\label{line:tagcons}@

arch_cap_flags = caml_alloc_small(1, tag);
Store_field(arch_cap_flags, 0, arch_obj); /* BUG: NULL pointer reachable from OCaml value */

CAMLreturn(arch_cap_flags);
\end{lstlisting}
\end{minipage}

\Cref{lst:tagcons,line:tagcons} is a misunderstanding: \verb|Tag_cons| is meant to be used as the tag argument to \verb|caml_alloc*| for allocating a list element.
An empty list should instead use \verb|Val_emptylist|.

The effect of storing \verb|Tag_cons| (which is defined as the value \verb|0|) is the naked \verb|NULL| pointer. This wouldn't be valid in OCaml 5, but even for earlier OCaml versions attempting to iterate on this list would lead to an immediate segmentation fault.
Currently, the \verb|x86_physinfo_cap_flag| type is currently entirely abstract and unused, so there is no code that would iterate on this list currently, but the bug should nevertheless be fixed to avoid it being taken as an example when developing other code.

A full static analyser for detecting \glspl{nakedp} would be nice to have but is a larger project (the compiler provides a runtime checker~\cite{nnpchecker}), however minimal support can be added quite easily for detecting \glspl{nakedp} that are known constants (e.g.\ through constant propagation or value range analysis), see~\cref{lst:extend}.

\begin{minipage}{0.95\linewidth}
\begin{lstlisting}[caption={Minimal constant naked pointer detection},label=lst:extend,language={[Objective]Caml}]
(** [is_known_naked_pointer ctx rval]

  @returns [true] when [rval] is known to always be a naked pointer.
  If it is never a naked pointer, or it is unknown whether it is a naked pointer or not
  this returns [false].
*)
let is_known_naked_pointer ctx (rval: exp) =
  (* ask existing Goblint value range analysis whether a value is known for the RHS expression *)
  match ctx.ask (Queries.EvalInt rval) |> Queries.ID.to_int with
  | None -> false (* value not statically known *)
  | Some int ->
    (* integers and pointers are differentiated in OCaml through the LSB:
       pointers are always aligned to at least 2 bytes and have a LSB equal to 0.
       integers sacrifice one bit, are shifted left with low order bit always set to 1.
     *)
    Int64.equal 0L @@ Int64.logand 1L (IntOps.BigIntOps.to_int64 int)

(** [assign ctx lval rval]  computes the analysis state after an [lval = rval] statement. *)
let assign ctx (lval:lval) (rval:exp) =
  let typ = Cil.typeOfLval lval in (* type of LHS *)
  let () = if is_ocaml_value_type typ  then (* is the type [value]? *)
    if is_known_naked_pointer ctx rval then (* is it always a naked pointer? *)
      Messages.error (* we found a bug: naked pointers are no longer supported in OCaml 5 *)
        ~category:Messages.Category.(Cast TypeMismatch)
		"Naked pointer detected"
  in
  ctx.local
\end{lstlisting}
\end{minipage}

This cannot prove the absence of \glspl{nakedp}, but should have a low false positive rate because it only reports errors on constants that are stored in \verb|value| types that are not integers,
and not values allocated by the OCaml runtime.

\section{Evaluation}

Initially developed and tested on the Xen and XAPI
\footnote{The XAPI toolstack is developed by the XAPI project: a subproject of the Linux Foundation Xen Project. It is used in the commercial XenServer~\cite{xenserver} distribution. XenServer was initially developed at XenSource,~Inc.\ acquired by Citrix~Systems,~Inc.\ in 2007, which was merged into Cloud~Software~Group,~Inc.\ in 2022 and is currently a separate Business Unit. Cloud~SG~UK~Ltd.\ is Cloud~Software~Group's UK subsidiary.} project source code.
It runs as a GitHub Action, using Goblint's existing support for producing a SARIF~\cite{githubsarif} file.
When any errors are detected it will highlight the source code location with the diagnostic
message generated by the tool in the GitHub UI automatically.
Currently, has 0 false positives and all reported bugs got fixed.
Building the static analyser, generating the primitive model, and running the static analyser takes about 12 m,
analysing \~100 C primitives.

May test on a wider selection of OCaml packages by ICFP\@.

\subsection{Limitations}\label{sec:lim2}

See also~\cref{sec:lim2}.

Fully verifying the correctness of C code linked with OCaml is beyond the scope of this tool, however verifying the safety of the C ``glue code'' in isolation is still useful:
\begin{itemize}
	\item the correctness of the C library can be verified by other static analysis tools, bounded model checking tools, or instrumented at runtime with various sanitizers (or may not even be written in C)
	\item the correctness of OCaml code is ensured by OCaml's type system (as long as escape hatches like \verb|Obj| are not used)
	\item the correctness of the OCaml to C glue code would be the responsibility of this tool, if this glue code is kept at a minimum
\end{itemize}

To verify the correctness of the glue code we need the following:
\begin{itemize}
	\item absence of undefined behaviour in the C code itself. Some built-in verification methods in Goblint can help here. We need this because otherwise even basic properties like code reachability or value range analysis can be invalidated with a simple stray pointer write.
	\item precision of the analysis. Goblint can warn when it doesn't have enough information or when an analysis is unsound. Unfortunately this can also lead to a high number of false positives.
	\item knowledge about the shape of the received and constructed OCaml types. The OCaml \verb|cmt| file would have this information, but it is currently not taken advantage of. E.g.\ we need to know whether we should expect an integer, a list, or a tree.
	\item \gls{GC} safety properties. Documented in the manual, e.g.\ don't dereference OCaml values when the runtime/domain lock is not held.
	\item the lifetime of OCaml values. Would need to know about OCaml \gls{GC} root registrations, the interactions with callbacks can be quite complex, but necessary (e.g.\ if we want to pass an OCaml closure to invoke from a C callback, and some additional data to pass from that callback). This becomes even more complex is \verb|ctypes| is used because then the OCaml value we received may not be alive if the \gls{GC} executes, which can be a problem if we received a pointer managed by an OCaml value, such as a memory region, but the parent value is now no longer considered ``alive''.
	\item model where callbacks can be invoked. Invoking them directly in the C glue code is fine, but if they get invoked due to a callback from a C library, then we'd have to model/analyse the C library itself to know when it might call us, and it wouldn't be practical to do that, this can only be approximated (which may lead to false positives).
\end{itemize}

Although creating an analyser that works for all possible C bindings would be very difficult, a different approach is proposed here: get the analyser working on one project at a time, and make it able to analyse the code patterns used within that project. Usually a project will have a limited number of code patterns/styles used for C bindings.
However, this doesn't mean that we add detection for just one bug at a time making the tool very specialized and hard to extend. We propose to teach the tool one concept at a time guided by the kind of bugs that are most often seen in a project.

E.g.\ check the runtime lock safety property for all OCaml values and OCaml runtime calls (the focus of the current paper), and a future improvement that introduces shape analysis could start with teaching it about modelling the shape of OCaml lists (and other data types with a similar structure). Then the analysis could show a warning when it detects that a value must always be a list, but that list is not always constructed correctly (e.g.\ if you store a non-zero integer or a string inside it).

\section{Related and Future Work}\label{section:related}

This tool is also presented at the Xen project's annual summit~\cite{nogremlins}, however that presentation
is focused on how to use this for Xen, and the specifics of how the OCaml \gls{GC} works for the kind of bindings Xen has.
This presentation is more general on how to use the tool on the wider ecosystem of OCaml packages, and how it works internally.

The tool will be released in 2 phases: first the small header generator tool from~\cref{sec:arity} will be release on opam.
This tool has 0 build dependencies for running the tool (it of course has test dependencies).
Then the static analyser will be made available on opam once the author has done some wider testing on packages in the opam ecosystem.
It is the author's hope to complete this testing, and provide some usage instructions by the time of the workshop, however it should be available in an experimental state in a GitHub repository.

Auto-generating C stubs using~\cite{ctypes} is another approach. However, in the author's experience this only works
reliably when all values can be copied or passed by value. Strings lifetimes are difficult to get correct,
and in fact due to a bug it was not possible to correctly use them on old versions of ctypes.
This has been reported on the ctypes mailing list with a self-contained reproduction, and the buggy binding rewritten by hand.
The rewritten binding worked reliably ever since, while the original one corrupted the string periodically during automated testing.
Retesting and fixing this would be part of future work.

There are a lot of C\footnote{when interfacing with safe languages there are other alternatives~\cite{ml2018ocamlsaferust,bour2019camlroot,munchmaccagnoni:hal-03910313,ocaml-interop}.}~static analysers
 (some even written in OCaml), but to the author's knowledge none of them
know about OCaml's \gls{GC} safety rules. In theory, it may be possible to use the auto-generated+handwritten OCaml runtime model
with other static analysers though.

There was a tool called ``Saffire''~\cite{foster2007improving,10.1145/1377492.1377493,saffireweb}, but unfortunately it is no longer being developed.
It would be useful to incorporate the shape analysis approaches used by it into \verb|lintcstubs| in the future.

\section{Future Work}

The header generator for C stubs in~\cref{sec:arity} should be part of the build system by default:
initially as a tool available on \verb|opam| that users can use via custom dune rules,
and if there is interest then as part of dune, and eventually part of the compiler as a standalone tool or flag for \verb|ocamlc|
(it is a very small tool).
It should be possible to do this in a backwards compatible way, e.g.~\cref{lst:generated}

\begin{lstlisting}[label=lst:generated,caption={Backwards compatible header inclusion},language={C}]
#include <caml/mlvalues.h>
#ifdef OCAML_HAS_GENERATED_HEADER
#include "mystub.h"
#endif
\end{lstlisting}

Currently, the tool has only been tested on Xen, the XAPI project and the OCaml standard library itself.
By the time of the conference the author hopes to have results on more packages,
and the tool should be available as open-source part of the XAPI project.

Now that we have a tool that can check some safety properties automatically it would be interesting to run this on the
ctypes generated code, try and reproduce the original bug and find the root cause of that bug and fix it (if still present,
ctypes has evolved a lot). However, that will be part of future work.

\begin{acks}
The author would like to thank both the anonymous reviewers and colleagues for their constructive feedback on the paper.
\end{acks}

\printglossaries{}
\printbibliography{}
\end{document}